\documentclass[a4paper,11pt]{article}
\pdfoutput=1

\usepackage{jinstpub}

\usepackage{graphicx}
\usepackage{amsmath,esint}
\usepackage{subfig}
\usepackage{enumitem}
\usepackage{setspace}

\usepackage{caption}
\title{Advances in IceCube ice modelling \newline \& what to expect from the Upgrade}

\author[a,1]{M. Rongen,\note{Corresponding author.}}
\author[b]{D. Chirkin}

\affiliation[a]{Institute of Physics, University of Mainz, D-55099 Mainz, Germany}
\affiliation[b]{Dept. of Physics and WIPAC, University of Wisconsin, Madison, WI 53706, USA}

\emailAdd{mrongen@uni-mainz.de}

\abstract{The IceCube Neutrino Observatory instruments about 1 km$^3$ of deep, glacial ice at the geographic South Pole using 5160 photomultipliers to detect Cherenkov light from relativistic, charged particles. Most IceCube science goals rely on precise understanding and modelling of the optical properties of the instrumented ice. A peculiar light propagation effect observed by IceCube is an anisotropic attenuation, which is aligned with the local flow of the ice. Recent efforts have shown this effect is most likely due to curved photon trajectories resulting from the asymmetric light diffusion in the birefringent polycrystalline microstructure of the ice. This new model can be optimized by adjusting the average orientation, size and shape of the ice crystals. We present the parametrization of the birefringence effect in our photon propagation simulation, the fitting procedures and results. The anticipated potential of calibration instrumentation in the upcoming IceCube Upgrade to improve on known shortcomings of the current ice modelling is also discussed.}

\arxivnumber{2108.03291} % only if you have one

\collaboration[c]{on behalf of the IceCube Collaboration$^*$\note[*]{Full author list and acknowledgments are available at \href{https://icecube.wisc.edu/collaboration/authors/\#collab=IceCube&date=2021-05-18&formatting=web&tag=VLVnT+2021}{icecube.wisc.edu}.}}

\proceeding{9$^{\text{th}}$ Very Large Volume Neutrino Telescope Workshop (VLVnT-2021)\\
  18-21 May 2021\\
  Valencia, Spain}

\begin{document}
\toccontinuoustrue
\notoc
\maketitle
\flushbottom

\section{Introduction}

\begin{figure}
\centering
\begin{minipage}{.63\textwidth}
    \centering
    \includegraphics[width=0.99\textwidth]{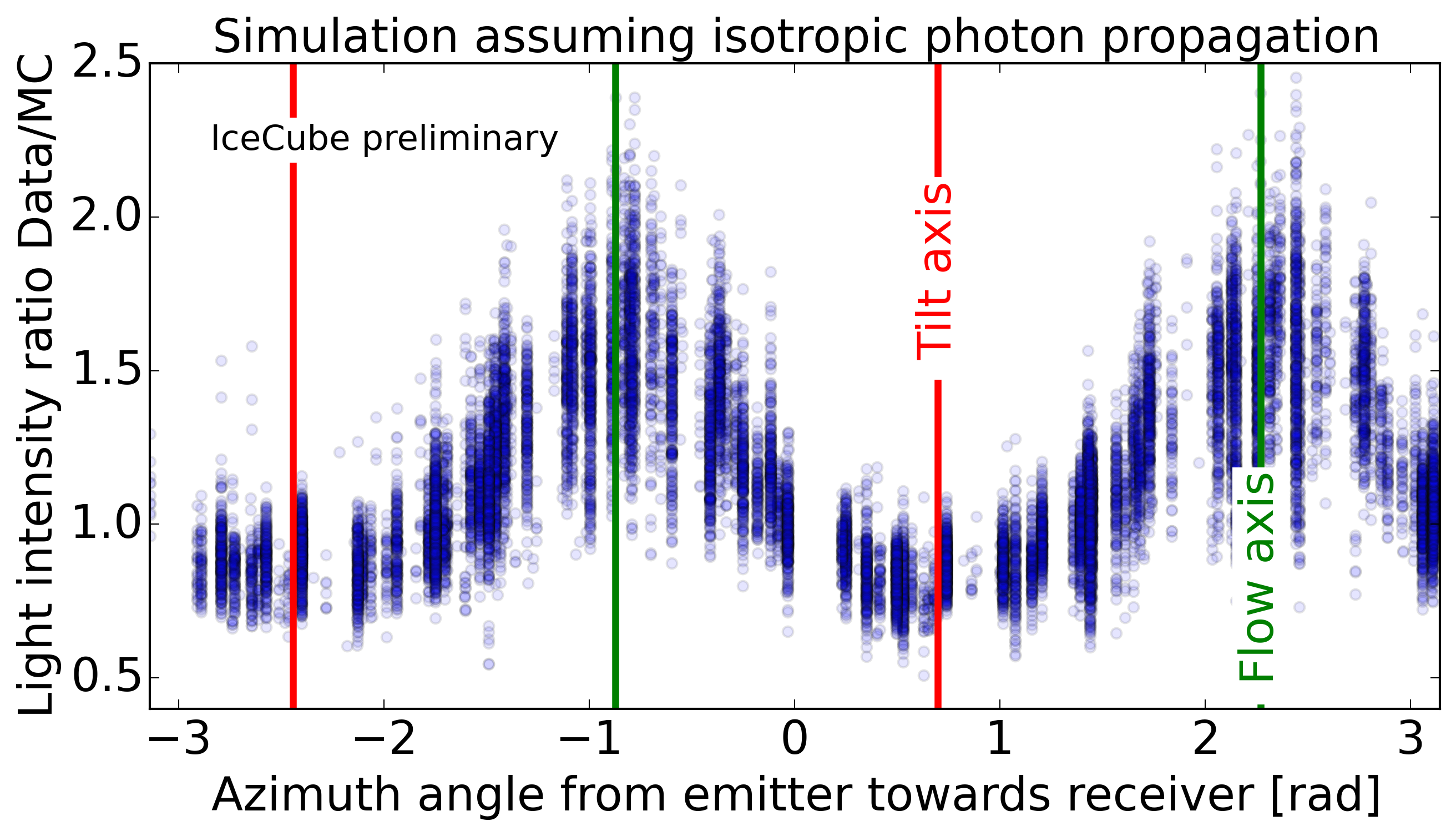}
    \caption{Optical ice anisotropy seen as azimuth dependent intensity excess in flasher data.}
    \label{fig:ratio}
\end{minipage} \hfill % 
\begin{minipage}{.34\textwidth}
    \centering
    \includegraphics[width=0.99\textwidth]{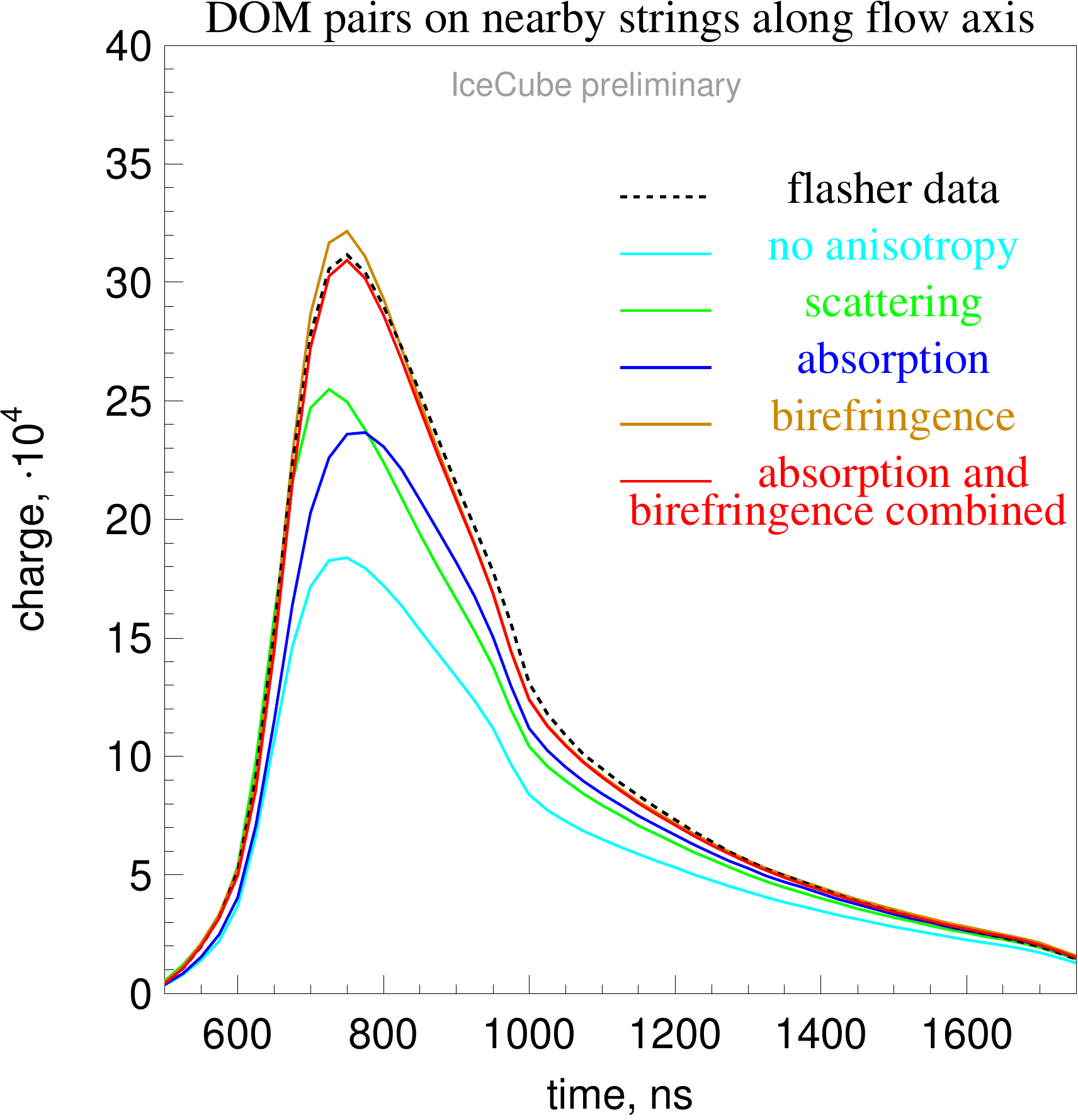}\hfill
    \caption{Comparison of fit quality achieved with different models of anisotropy.}
    \label{fig:lightcurves}
\end{minipage}
\end{figure}

The IceCube Neutrino Observatory is a cubic-kilometer Cherenkov detector instrumenting depths between 1450\,m and 2450\,m in the ice at the geographic South Pole \cite{detector:paper}. The optical properties of the ice were determined with a measurement as described in \cite{Aartsen2013} using flasher LEDs integrated into each Digital Optical Module (DOM).
As previously reported \cite{ICRC13_anisotropy,TC:logger}, the ice  exhibits a strong anisotropy in light propagation, aligned with the local ice flow direction. Measured at $\sim$125\,m from an isotropic emitter (averaging over many flashers), about twice as much light reaches DOMs on the flow axis then on the orthogonal (tilt) axis (see Figure \ref{fig:ratio}). At the same time the arrival time distributions are nearly unchanged compared to a simulation expectation without anisotropy.
Several parametrizations directionally modifying the scattering function, absorption or scattering coefficients  have been explored in the past with some success. However, none of them were able to fit intensity and timing distributions simultaneously. Departing from the paradigm that the scattering of light in ice is only caused by particulate impurities, light deflection resulting from asymmetric diffusion in birefringent ice polycrystals with preferential c-axes distributions was proposed in 2019 \cite{ICRC19:anisotropy}.  Building on this work, we present an IceCube ice model fitted to LED data, called SpiceBFR.

\section{Simulating diffusion patterns}

Simulating each crystal crossing over the full scale required during photon propagation is computationally unfeasible using the software introduced in \cite{ICRC19:anisotropy}. Thus ensembles of photon positions and directions for a given initial photon direction and crystal realization in otherwise optically perfect ice, called diffusion patterns, are simulated for 1000 crystal crossings ($\sim$1\,m). These diffusion patterns are then parametrized in terms of their diffusion, deflection and displacement (see section \ref{section:parametrization}) and applied in photon propagation (section \ref{section:implementation}). 
Considering Snell's law, both the encountered refractive indices as well as the boundary surface orientations dictate the resulting diffusion. The refractive index experienced by the extraordinary rays depends on the opening angle between the wave vector and the crystal axis of the traversed grain. Thus the distribution of c-axes found in many crystals, also referred to as fabric, needs to be modeled and sampled from for each simulated grain. This is realized as described in \cite{TC:caxes}, with the characteristics fabric parameters $\ln(S_1/S_2)$ and $\ln(S_2/S_3)$ as also used and measured during ice core analysis.
Since the average grain shape is non-spherical, the distribution of boundary surface (or face) orientations depends on photon direction. Assuming the randomly shaped, but on average elongated, ice crystal polyhedra are (on average) approximated well by an ellipsoid, the face orientation distribution can be sampled from analytic functions  \cite[p.~169]{Rongen:PHD}. We here restrict ourselves to spheroids instead of arbitrary ellipsoids, as preferred by early fits.

\section{Parametrizing diffusion patterns}
\label{section:parametrization}

\begin{figure}
\centering
\begin{minipage}{.44\textwidth}
    \centering
    \includegraphics[width=\linewidth]{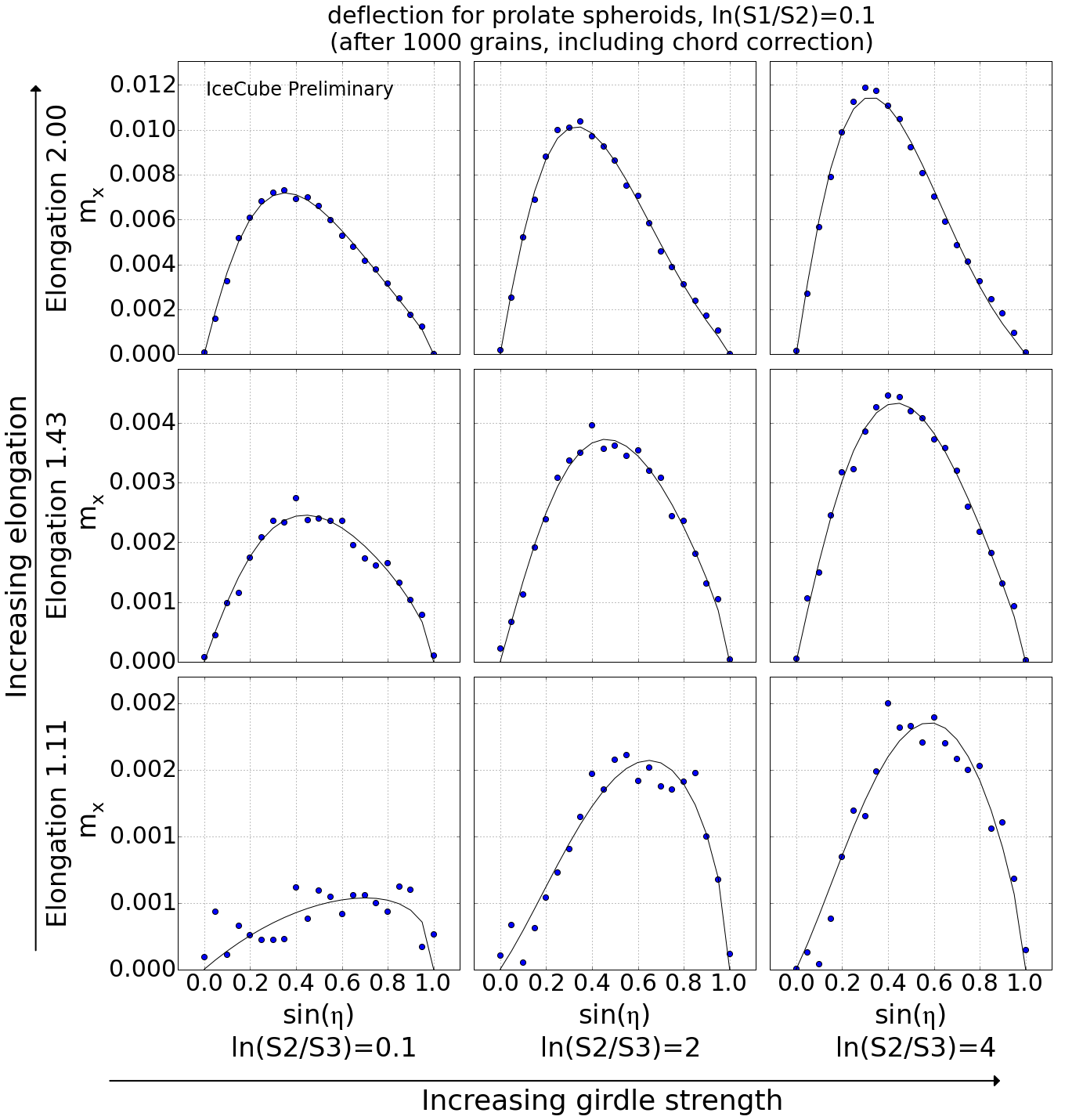}
    \caption{Deflection vs. opening angle to the flow for nine crystal configurations.}
    \label{fig:deflections}
\end{minipage} \hfill % 
\begin{minipage}{.52\textwidth}
    \centering
    \includegraphics[width=\linewidth]{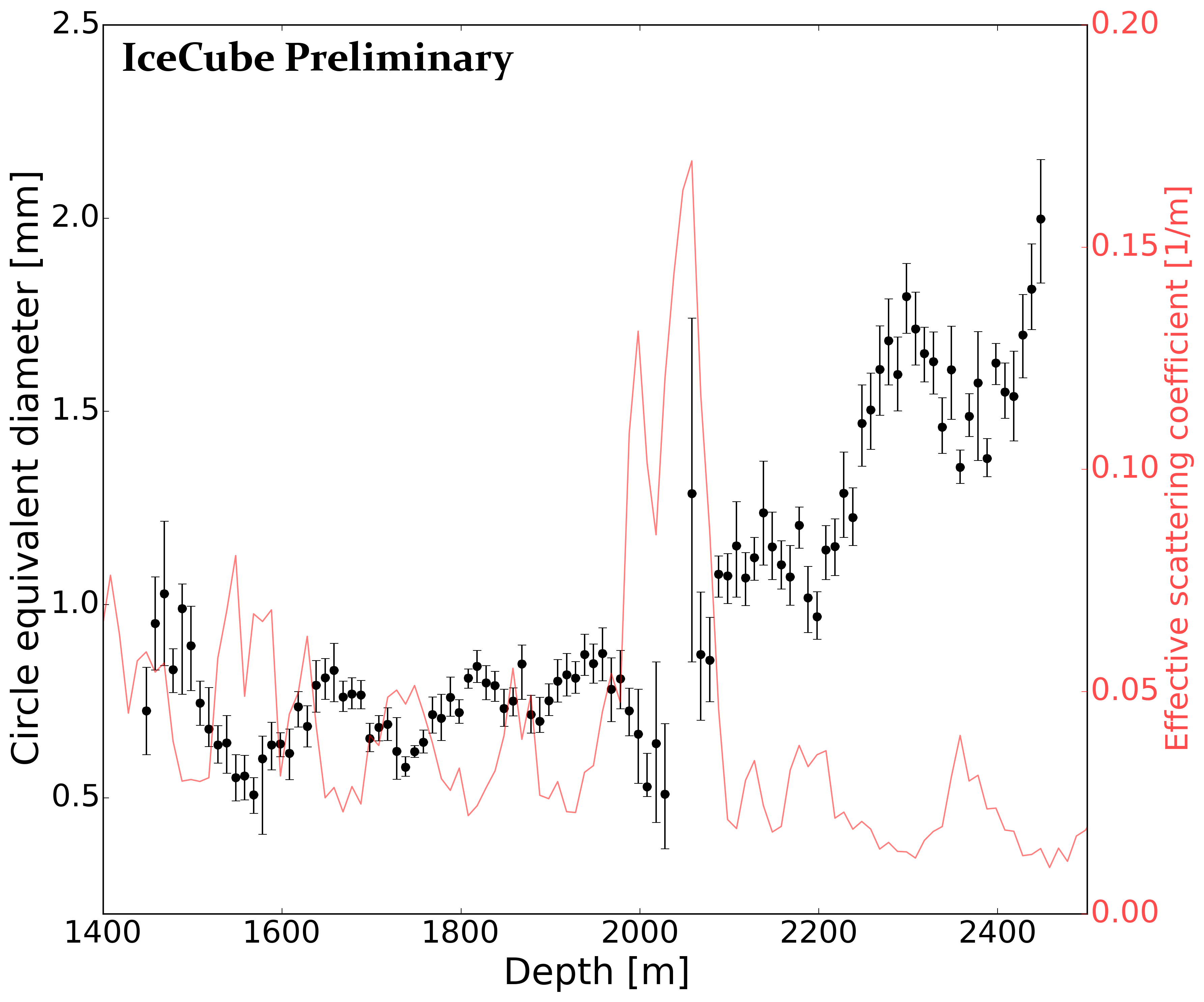} 
    \caption{Best fit crystal sizes. Error bars denote the statistical uncertainty only.}
    \label{fig:size}
\end{minipage}
\end{figure}

Diffusion patterns (examples can be found in \cite{ICRC19:anisotropy}) have been simulated for a wide range of spheroid elongations and fabric parameters. These diffusion patterns have a strong central core with a tail dominated by mainly single large-angle reflections. The distribution is modeled as a 2d-Gaussian on a sphere, allowing scaling (with distance) relationships for mean displacement and width. 
The three parameters of the 2d-Gaussian on a sphere are the two widths (in the directions towards the flow, $\sigma_x$, and perpendicular to it, $\sigma_y$), and a single mean deflection towards the flow, $m_x$. Mean deflection in the perpendicular direction was zero for all considered cases. Because we mainly simulate small deflections, we simulated the 2d-Gaussian in Cartesian coordinates, and then projected that to the sphere with an inverse stereographic projection. The three parameters were fitted to the following functions of angle $\eta$ of initial photon direction with respect to the ice flow, for simulations with a fixed number of 1000 crystal crossings:
\begin{eqnarray}
    m_x = \alpha \cdot \arctan(\delta \cdot \sin \eta \cos \eta) \cdot \exp(-\beta \sin \eta + \gamma \cos \eta) \quad (\mbox{deflection})\\
    \sigma_{x,y} = A_{x,y} \cdot \exp(-B_{x,y} \cdot (\arctan(D_{x,y} \sin \eta))^{\mbox{$C_{x,y}$}}) \quad (\mbox{diffusion}).
\end{eqnarray}
These functions were found to describe all considered crystal realizations with only 12 free parameters ($A_{x,y}$..$D_{x,y}$ \& $\alpha$..$\delta$). Figure \ref{fig:deflections} shows the mean deflection for nine crystal configurations. Note that increasing elongation has a stronger effect compared to a strengthening fabric. 

\section{Applying diffusion patterns in photon propagation}
\label{section:implementation}

During photon propagation directions are only updated upon scattering. To minimize the computational burden, the new birefringence anisotropy is discretized and also evaluated only at the scattering sites. This requires scaling the diffusion \& deflection to the number of traversed grains between two scatterings. This introduces a new model parameter, the average grain size, and requires taking into account the different average crystal chord lengths as a function of propagation direction (described in \cite{Rongen:PHD}). As expected in a diffusion process, the deflection scales linear with the number of traversed grains ($m_x\propto n$) and the diffusion as the square root ($\sigma_{x(y)}\propto n$). To decouple the fitting of anisotropy properties from the overall ice diffusion / scattering strength as much as possible, the effective scattering Mie coefficient was reduced by the amount resulting from the birefringence induced light diffusion. Updating not only a photon's direction, but also the photon coordinates (as it shifts transversely with respect to straight-path expectation), further improves the quality of description of data and was implemented as described in \cite{CHEECKYICRCBFR}.

\section{Fitting to flasher data} \label{section:fitting}

Using the described model, 4 parameters are required to specify a birefringence anisotropy realization. Allowing for a correction to the Mie coefficients adds a further two parameters. As minimizing that many parameters for all layers in the ice model is not computationally feasible, we need to identify some which are either depth independent or have a small effect.
The required pre-fits, as well as the final depth evaluation, were performed following the method described in \cite{Aartsen2013}. We minimize the summed LLH\footnote{The minus log-likelihood, denoted here as LLH, is akin to the saturated Poisson likelihood, and can similarly be used as a measure of the goodness-of-fit \cite{llh}.} which compares the {\it single-LED} data set (where all 12 LEDs on all in-ice DOMs were flashed one at a time) with full photon propagation simulation of these events taking into account precisely known DOM orientations as measured in \cite{Dima2021}. 
Fits for individual layers were carried out by only including LEDs situated within the considered ice layer into the LLH summation. 
During the pre-fits the following behavior was noted:
Given a girdle fabric ($\ln(S_1/S_2) >> \ln(S_2/S_3)$), the actual fabric strength has a small effect and cannot be distinguished by the data. Accordingly the fabric has been fixed to values measured in the deepest sections of the South Pole Ice Core, SPC14, \cite{SPC14:caxis} ($\ln(S_1/S_2)=0.1 \,\&\, \ln(S_2/S_3)=4$).
The fit is largely degenerate in crystal elongation and size. Thus, the elongation was fixed to 1.4, which is a good fit at all layers and similar to the values measured in SPC14 \cite{Alley2021}.
Fitting the remaining parameters, crystal size and absorption \& scattering correction for all layers, yields a significant improvement as seen for example in the average light curves in Figure \ref{fig:lightcurves} (birefringence only line). Still the best-fit does not perfectly match the data and the required crystal sizes are far smaller than expected from SPC14. 
After thorough checks, we decided to reintroduce scattering as well as absorption anisotropy, both following the formalism as described in \cite{ICRC13_anisotropy}, into the fit. The fit does not make use of the scattering anisotropy, but surprisingly the absorption anistropy is mixed into the birefringence model with a significant non-zero contribution. This means a departure from a first-principle model, but was adopted for its improvement in data-MC agreement. After including the absorption anisotropy, crystal size and absorption \& scattering correction were again fitted for all layers. Figure \ref{fig:size} depicts the best fit stratigraphy of grain sizes. The overall grain size of $\sim$1\,mm as well as the increase in older and cleaner ice are as generally expected in glaciology \cite{EPICA2004, Alley2021}.  As seen in Figure \ref{fig:lightcurves} the new model significantly improves in matching the flasher data both in terms of timing and total intensity with regards to older models and for the first time achieves an excellent data-MC agreement.

\section{Future studies using the Pencil Beam in the IceCube Upgrade}
\label{Upgrade}

\begin{figure}
\centering
\begin{minipage}{.49\textwidth}
    \centering
    \includegraphics[width=0.9\linewidth]{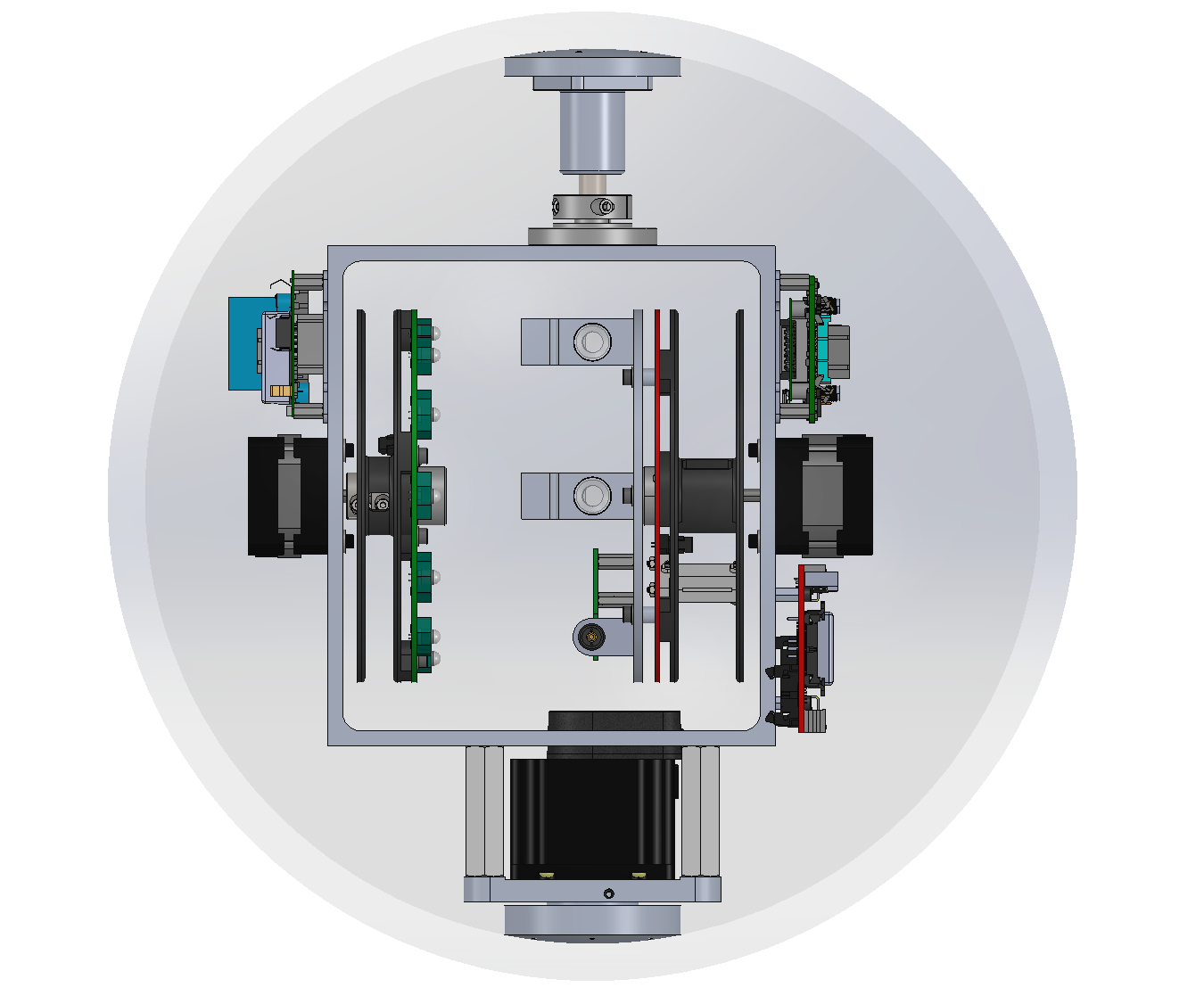}
    \caption[font=small]{Design sketch of the Pencil Beam.}
    \label{fig:PB}
\end{minipage} \hfill % 
\begin{minipage}{.49\textwidth}
    \centering
    \includegraphics[width=0.9\linewidth]{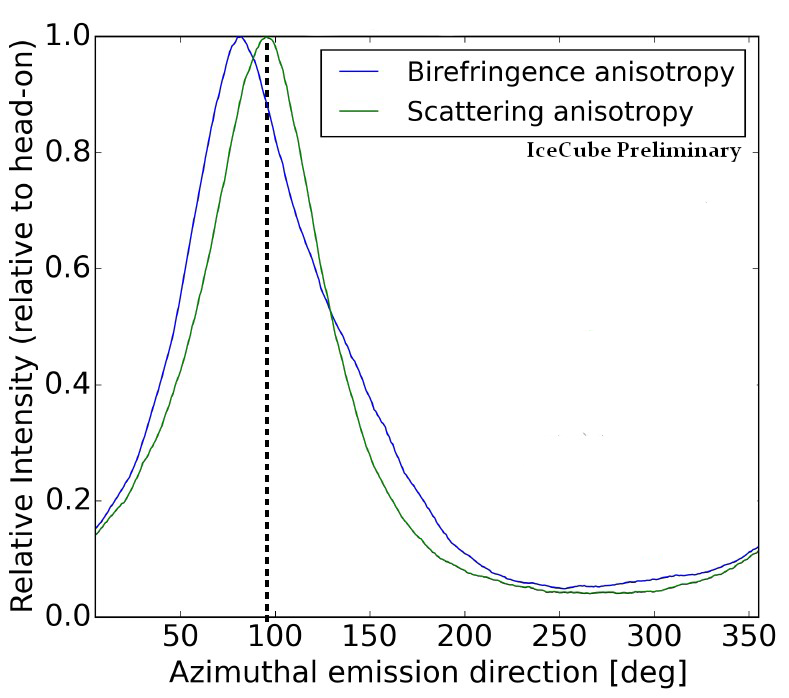} 
    \caption[font=small]{Intensity profile as the PB sweeps over a receiver given two different anisotropy models.}
    \label{fig:PBsim}
\end{minipage}
\end{figure}

The IceCube Upgrade \cite{IcecubeUpgrade}, planned to be deployed during the 2022/23 season, marks the first extension of the IceCube Detector. Over 700 additional modules, including a number of stand-alone calibration devices\cite{POCAM}, will be deployed on seven additional strings. Of particular interest for the anisotropy are eleven so called Pencil Beam (PB) devices, as depicted in Figure \ref{fig:PB}. They allow for a laser-like beam to be directed in arbitrary directions, enabling in particular sweeps over receiver directions (see Figure \ref{fig:PBsim}). The birefringence induced deflection yields a unique signature, where the emission direction of maximum received intensity is offset from the geometric direction of the receiver. Measuring sweeping profiles for several emitter-receiver pairs at different orientations will allow to disentangle absorption and birefringence contributions to the anisotropy at high precision.

\section{Summary and Outlook}
\label{section:summary}

A model combining anisotropic absorption with light deflection resulting from propagation through the birefringent ice polycrystal significantly improves on previous ice models. The model yields a near perfect data-MC agreement for flasher data in timing and intensity variables and will improve on event reconstructions while reducing systematic biases.  
In the fitting process the average crystal size in the detector is deduced. While the birefringence model has been deduced from first principle, the absorption contribution is so far unmotivated. Disentangling the absorption and birefringence contributions will be the focus of future studies, in particular in the IceCube Upgrade.

\bibliographystyle{JHEP}
\bibliography{skeleton}

\providecommand{\href}[2]{#2}\begingroup\raggedright\begin{thebibliography}{10}

\bibitem{detector:paper}
{\scshape IceCube} collaboration, \emph{{The IceCube Neutrino Observatory:
  Instrumentation and Online Systems}},
  \href{https://doi.org/10.1088/1748-0221/12/03/P03012}{\emph{JINST} {\bfseries
  12} (2017) P03012} [\href{https://arxiv.org/abs/1612.05093}{{\ttfamily
  1612.05093}}].

\bibitem{Aartsen2013}
{\scshape IceCube} collaboration, \emph{Measurement of south pole ice
  transparency with the {IceCube} {LED} calibration system},
  \href{https://doi.org/10.1016/j.nima.2013.01.054}{\emph{Nucl. Instr. Meth.
  Phys. Res.} {\bfseries 711} (2013) 73}.

\bibitem{ICRC13_anisotropy}
{\scshape IceCube} collaboration, \emph{{Evidence of optical anisotropy of the
  South Pole ice}}, {\emph{ICRC} (2013) }
  [\href{https://arxiv.org/abs/1309.7010}{{\ttfamily 1309.7010}}].

\bibitem{TC:logger}
M.~Rongen, R.~C. Bay and S.~Blot, \emph{Observation of an optical anisotropy in
  the deep glacial ice at the geographic south pole using a laser dust logger},
  \href{https://doi.org/10.5194/tc-14-2537-2020}{\emph{The Cryosphere}
  {\bfseries 14} (2020) 2537}.

\bibitem{ICRC19:anisotropy}
{\scshape IceCube} collaboration, \emph{{Light diffusion in birefringent
  polycrystals and the IceCube ice anisotropy}},
  \href{https://doi.org/10.22323/1.358.0854}{\emph{PoS} {\bfseries ICRC2019}
  (2020) 854} [\href{https://arxiv.org/abs/1908.07608}{{\ttfamily
  1908.07608}}].

\bibitem{TC:caxes}
M.~Rongen, \emph{Brief communication: Sampling c-axes distributions from the
  eigenvalues of ice fabric orientation tensors},
  \href{https://doi.org/10.5194/tc-2019-204}{\emph{The Cryosphere Discussions}
  {\bfseries 2019} (2019) }.

\bibitem{Rongen:PHD}
M.~Rongen, \emph{{C}alibration of the {I}ce{C}ube neutrino observatory},
  dissertation, RWTH Aachen University, Aachen, 2019.
\newblock 10.18154/RWTH-2019-09941.

\bibitem{CHEECKYICRCBFR}
{\scshape IceCube} collaboration, \emph{{A novel microstructure-based model to
  explain the IceCube ice anisotropy}},
  \href{https://doi.org/10.22323/1.395.1119}{\emph{PoS} {\bfseries
  395(ICRC2021)} (2021) } [\href{https://arxiv.org/abs/2107.08692}{{\ttfamily
  2107.08692}}].

\bibitem{llh}
D.~Chirkin, \emph{{Likelihood description for comparing data with simulation of
  limited statistics}},  \href{https://arxiv.org/abs/1304.0735}{{\ttfamily
  1304.0735}}.

\bibitem{Dima2021}
{\scshape IceCube} collaboration, \emph{{A calibration study of local ice and
  optical sensor properties in IceCube}},
  \href{https://doi.org/10.22323/1.395.1023}{\emph{PoS} {\bfseries
  (ICRC2021)1023} }.

\bibitem{SPC14:caxis}
{Donald E., Voigt, }, \emph{{c-Axis Fabric of the South Pole Ice Core, SPC14}},
  \href{https://doi.org/10.15784/601057}{\emph{{U.S. Antarctic Program (USAP)
  Data Center}} (2017) }.

\bibitem{Alley2021}
R.~Alley et~al., ``Physical properties of the south pole ice core, spc14.''
  {IceCube Polar Science Workshop}, 2021.
\newblock https://events.icecube.wisc.edu/event/128/contributions/7313/.

\bibitem{EPICA2004}
A.~Laurent et~al., \emph{Eight glacial cycles from an antarctic ice core},
  \href{https://doi.org/10.1038/nature02599}{\emph{Nature} {\bfseries 429}
  (2004) 623}.

\bibitem{IcecubeUpgrade}
{\scshape IceCube} collaboration, \emph{{The IceCube Upgrade - Design and
  Science Goals}}, \href{https://doi.org/10.22323/1.358.1031}{\emph{PoS}
  {\bfseries ICRC2019} (2021) 1031}
  [\href{https://arxiv.org/abs/1908.09441}{{\ttfamily 1908.09441}}].

\bibitem{POCAM}
F.~Henningsen et~al., \emph{{A self-monitoring precision calibration light
  source for large-volume neutrino telescopes}},
  \href{https://doi.org/10.1088/1748-0221/15/07/P07031}{\emph{JINST} {\bfseries
  15} (2020) P07031} [\href{https://arxiv.org/abs/2005.00778}{{\ttfamily
  2005.00778}}].

\end{thebibliography}\endgroup

\end{document}